\documentclass[twocolumn,runningheads,natbib]{svj2mod}
\smartqed  
\usepackage{graphicx}
\usepackage{amssymb}
\journalname{\textit{Astrophysics and Space Science} (2007)
 \textbf{308}: 353--361 / 
{\bf Corrected electronic version: September 4, 2007}}

\begin{document}

\title{Heat Blanketing Envelopes and Thermal Radiation
 of Strongly Magnetized Neutron Stars \thanks{Work 
 supported in parts by
 RFBR (grants 05-02-16245 and 05-02-22003),
 FASI (grant NSh-9879.2006.2), and CNRS French-Russian program
 (grant PICS 3202)}
}


\author{Alexander Y. Potekhin       \and
        Gilles Chabrier   \and
        Dmitry G. Yakovlev
}

\authorrunning{A.~Y. Potekhin, G. Chabrier, \& D.~G. Yakovlev}

\institute{A. Y. Potekhin \at
              Ioffe Physico-Technical Institute, Politekhnicheskaya 26,
              194021 St.\,Petersburg, Russia \\
              \email{palex@astro.ioffe.ru}           
           \and
           G. Chabrier \at
              Ecole Normale Sup\'erieure de Lyon, CRAL (UMR 5574 CNRS),
              46 all\'ee d'Italie, 69364 Lyon, France \\
              \email{chabrier@ens-lyon.fr}           
           \and
           D. G. Yakovlev  \at
              Ioffe Physico-Technical Institute, Politekhnicheskaya 26,
              194021 St.\,Petersburg, Russia \\
              \email{yak@astro.ioffe.ru}           
}

\date{Received: July 5, 2006 / Accepted: October 31, 2006}

\maketitle

\begin{abstract}
Strong ($B \gg 10^9$ G) and superstrong ($B \gtrsim 10^{14}$ G)
magnetic fields profoundly affect many thermodynamic and
kinetic characteristics of dense plasmas in neutron star envelopes. 
In particular, they produce strongly anisotropic 
thermal conductivity in the
neutron star crust and 
modify the equation of state and radiative opacities  in
the atmosphere, which are major ingredients of the cooling
theory and spectral atmosphere models. As a result, both the
radiation spectrum and the thermal luminosity of a neutron star can be
affected by the magnetic field. We briefly review these effects
and demonstrate the influence of magnetic
field strength on the thermal structure of an isolated  neutron star,
putting emphasis on the differences brought about 
by the superstrong fields and high temperatures of magnetars.
For the latter objects,
it is important to take proper account of a combined effect
of the magnetic field on thermal conduction
and neutrino emission  at densities $\rho\gtrsim10^{10}$
g~cm$^{-3}$. We show that
the neutrino emission puts a $B$-dependent upper limit
on the effective surface temperature of a cooling neutron star.
\end{abstract}

\section{Introduction}
\label{sect:intro}
Thermal emission from neutron stars 
can 
be used to measure 
the magnetic field, temperature, and composition of neutron-star envelopes,
and to constrain the properties of matter 
under extreme conditions
(see, e.g., \citealt{YakP04,Yak-ea05}, and references therein).
To achieve these goals, one should use reliable
models of the atmosphere or condensed
surface, where the thermal spectrum is formed, and of deeper 
layers, 
which provide thermal insulation of hot stellar 
interiors.
In these layers, the effects of strong magnetic fields
can be important.
In recent years, significant progress has been achieved 
in the theoretical description of neutron-star envelopes 
with strong magnetic fields,
but new challenges are put forward by observations of magnetars.
In Sect.~\ref{sect:1} we briefly overview recent work
on the construction of models of neutron star atmospheres 
with strong magnetic fields and on the modeling of spectra
of thermal radiation formed in an atmosphere or
at a condensed surface.
We list important unsolved theoretical problems
which arise in this modeling.
In Sect.~\ref{sect:2}, after a brief review of the effects of strong
magnetic fields on the thermal structure and effective
temperature of neutron stars,
we describe our new calculations of the thermal structure.
Compared to the previous results \citep{PY01,PYCG03},
we have taken into account neutrino energy losses 
in the outer crust of the star. We show that neutrino emission
strongly affects the temperature profile in a
sufficiently hot
 neutron star
and 
places an upper limit on its surface
temperature $T_\mathrm{s}$ and photon luminosity $L_\gamma$.

\section{Thermal Emission from Magnetized Surface Layers of a Neutron Star}
\label{sect:1}

\subsection{Neutron Star Atmosphere}
\label{sect:atm}

It was realized long ago \citep{Pavlov-ea95}
that a neutron star
atmosphere model should properly include the effects 
of a strong magnetic field
and partial ionization.
Models of \emph{fully ionized} neutron star 
atmospheres with strong magnetic fields
were constructed by several research groups 
\citep[e.g.,][and references therein]{Shib92,Zane00,HoLai02}.
The most recent papers highlighted the effects that
can be important for atmospheres of magnetars: the ion cyclotron feature
\citep{HoLai,Zane01} and vacuum polarization, 
including a conversion of normal radiation modes propagating in the 
magnetized atmosphere \citep{HoLai02,LaiHo03}.

Early studies of \emph{partial ionization}
in the magnetized neutron star atmospheres 
(e.g., Rajagopal, Romani, \& Miller \citeyear{RRM}; reviewed by \citealt{ZP02}) 
were based on an oversimplified treatment of
atomic physics and nonideal plasma effects in strong magnetic fields.
At typical parameters, 
the effects of thermal motion of bound species are important.
So far these effects have been taken into account only for hydrogen plasmas.
Thermodynamic functions, 
absorption coefficients,
the dielectric tensor
and polarization vectors of normal radiation modes
in a strongly magnetized, partially ionized hydrogen plasma
have been obtained and used to calculate
radiative opacities 
and thermal radiation spectra \citep[see][and references therein]{KK}.

The summary of the magnetic hydrogen atmosphere models 
and the list of references is given by
\citet{Potekhin-SCCS}. The model is sufficiently reliable
at $10^{12}$~G $\lesssim B \lesssim 10^{13.5}$~G,
i.e., in the field range typical of isolated radio pulsars.
It provides realistic spectra of thermal X-ray radiation
\citep{KK}. 
\citet{PC04} extended this model to higher $B$.
However, there remain the following unsolved theoretical problems
that prevent to obtain
reliable results beyond the indicated field range.
\begin{itemize}
\item
 The calculated spectra  at
$B\gtrsim10^{14}$~G depend on the adopted model of mode 
conversion owing to
the vacuum resonance and on the description of 
the propagation of photons with
frequencies below the plasma frequency.  
Neither of these problems has been
definitely solved. Their solution is also  important for modeling the
low-frequency (UV and optical) tail of the spectrum.

\item
 At low $T$ or high $B$, hydrogen atoms recombine in
H$_n$ molecules and eventually form a condensed phase 
(see Sect.~\ref{sect:surf}).
Corresponding quan\-t\-um-mechanical data are very incomplete.

\item
 At $10^9\mbox{ G}\lesssim B \lesssim 10^{11}$~G, 
transition rates of moving H atoms have
 not been calculated because of their
complexity. 
There is the only one calculation of the energy spectrum
of bound states appropriate to this range of $B$
\citep{LozovikVolkov}.

\item
 A more rigorous treatment of radiative transfer in the atmosphere
requires solving the transfer equations for the Stokes parameters
which has not been done so far
for partially ionized atmospheres
(see, e.g., \citealt{LaiHo03,LaiAdel} for the cases 
of fully ionized atmospheres).
\end{itemize}

Finally, we note that it is still not possible to calculate accurate
atmospheric spectra at $B\gtrsim10^{12}$~G for chemical elements
other than hydrogen, because of the importance of
the effects of motion of atomic nuclei in the strong magnetic fields.
Apart from the H atom, these effects have been
calculated only for the He atom \citep{Hujaj03a,Hujaj03b},
which \emph{rests} as a whole,
but has a moving nucleus, 
and for the He$^+$ ion (\citealt*{BPV}; Pavlov \& Bezchastnov \citeyear{PB05}).
The data of astrophysical relevance for He$^+$
are partly published and partly in preparation 
\citep[see][]{PB05}; one expects to have
a He/He$^+$ magnetic atmosphere model available in the near future.

\subsection{Condensed Surface and Thin Atmosphere}
\label{sect:surf}

The notion that an isolated magnetic neutron star has a 
condensed surface was first put forward by  
\citet{Ruderman71}, who considered the iron surface.
 \citet{LS97} and \citet{Lai-RMP}
studied the phase diagram of strongly magnetized hydrogen and
showed that, when the surface temperature
$T_\mathrm{s}$ falls below some critical value (dependent 
of $B$), the atmosphere can undergo a phase
transition into a condensed state.
A similar phase transition occurs for the equation
of state of partially ionized, nonideal, strongly magnetized 
hydrogen plasma, constructed by Potekhin, Chabrier, \& Saumon (\citeyear{PCS})
 for $B\lesssim10^{13.5}$~G
and extended by \citet{PC04} to the magnetar field strengths.
It is analogous to the ``plasma phase transition'' suggested in plasma
physics at $B=0$ (see, e.g., Chabrier, Saumon, \& Potekhin \citeyear{CSP} for discussion and references).
According to Pote\-kh\-in et al.\ (\citeyear{PCS}), 
the critical point for the phase transition in the hydrogen plasma
is located
at the density $\rho_c\approx 143 B_{12}^{1.18}$ g~cm$^{-3}$
and temperature $T_c\approx 3\times10^5 B_{12}^{0.39}$~K,
where $B_{12}=B/10^{12}$~G. At $T < T_c$ the density $\rho_\mathrm{cond}$
of the condensed phase increases up to a few times of $\rho_c$.
On the other hand, according to \citet{Lai-RMP},
the surface density of a condensed phase for heavy elements is 
$\rho_\mathrm{cond}\approx 560 A Z^{-0.6} B_{12}^{1.2}$ g~cm$^{-3}$,
where $Z$ and $A$ are the charge and mass numbers of the ions.
These two estimates of $\rho_\mathrm{cond}$
are in qualitative agreement.
\citet{Lai-RMP}
estimates the critical temperature of hydrogen as 
$T_c \lesssim 0.1 E_\mathrm{s} \lesssim 10^{5.5}B_{12}^{0.4}$~K
($E_\mathrm{s}$ being the cohesive energy),  also in agreement
with the above estimate.
\citet{Jones86} calculated the cohesive energy for Ne and Fe 
at $10^{12}$~G$\lesssim B \leq10^{13}$~G, using the density functional
   theory (DFT) and obtained
$E_\mathrm{s}\sim0.1B_{12}$ keV.
Recently, \citet{MedinLai06b} performed DFT calculations of the cohesive
energies for zero-pressure condensed
hydrogen, helium, carbon, and iron
at $10^{12}$~G $\leq B\lesssim10^{15}$~G.
For instance, they found that
the cohesive energy per carbon atom
ranges from $\sim50$ eV at $B=10^{12}$~G to 20 keV at $10^{15}$~G.
The cohesive energy per iron atom varies
from $\sim0.8$ keV at $B=10^{13}$~G to 33 keV at $10^{15}$~G.
These calculations suggest $T_c$
of the same order of magnitude as the above estimate for hydrogen.

Note that
the models of \citet{PCS} and \citet{LS97} 
are constructed in the framework of 
the ``chemical picture'' of plasmas, whose
validity near the plasma
phase transition can be questionable \citep{CSP}. Thus the position 
(and the very existence) of the condensed surface requires
further theoretical investigation and experimental or observational 
verification. Hopefully, this can be done
by analyzing observations of thermal emission from neutron stars.

The thermal emission from the magnetized surface was  
studied by \citet{Brinkmann}, Turolla, Zane, \& Drake 
(\citeyear{Turolla-ea}),
\citet*{perez-azorin}, and
\citet{surfem}. The spectrum exhibits 
modest deviations from blackbody across a wide energy range, 
and shows mild absorption features associated with the ion 
cyclotron frequency
(energy $\hbar\omega_{ci}=6.3 B_{12} Z/A$ eV)
 and the electron plasma frequency 
 (energy $\hbar\omega_p=28\sqrt{\rho Z/A}$ eV, where $\rho$
 is in g~cm$^{-3}$). However, the 
 predictions of the ion cyclotron 
feature and the spectrum at lower frequencies 
are not firm.
The uncertainty arises from motion of the ions 
in the electromagnetic field
around their equilibrium lattice positions.
Most of the models treat the ions as fixed (non-moving). Only
\citet{surfem} considered two limits of fixed and free ions.
In reality, however, the ions are neither fixed nor completely free
(see \citealt{surfem} for estimates of possible 
uncertainties).

In addition, the condensed surface of a neutron star
can be surrounded by a ``thin'' atmosphere, which is transparent to
X-rays, but optically thick at lower wavelengths.
Such a hypothesis has been first invoked by \citet*{Motch-ea03}
for explaining the spectrum of the isolated neutron star 
RX J0720.4$-$3125. Recently, the hydrogen atmosphere
model, described in Sect.~\ref{sect:atm},
together with the condensed surface emission model of \citet{surfem}
have been successfully used for fitting the spectrum
of the isolated neutron star RX J1856.5$-$3754 \citep{Ho-RXJ}
(assuming the atmosphere to be ``thin'' as defined above).

\section{Heat Transport through Magnetized Envelopes}
\label{sect:2}

\subsection{Overview of Previous Work}

The link between the magnetized atmosphere and stellar interior
is provided by a \textit{heat blanketing (insulating) envelope}.
The solution of heat transport problem relates the effective
surface temperature $T_\mathrm{s}$ to the temperature 
$T_\mathrm{b}$ at the inner boundary of the blanketing envelope.
Without a magnetic field, it is conventional 
to place the inner boundary 
at 
$\rho=\rho_\mathrm{b}=10^{10}$ g~cm$^{-3}$.
In this case, it can be treated in the quasi-Newtonian approximation
with fractional errors $\lesssim10^{-3}$ \citep*{GPE}.
A strong magnetic field, however, greatly affects heat 
transport and, consequently, the thermal structure of the envelope.
The thermal structure of neutron star envelopes with radial magnetic
fields (normal to the surface) was studied by \citet{kvr88} (also see
\citealt{kvr88} for references to earlier work). His principal
conclusion was that the field reduces the thermal insulation of the
heat blanketing envelope due to the Landau quantization of electron motion.
The thermal structure of the envelope with magnetic fields
normal and tangential to the surface was analyzed by
\citet{hern85} and \citet{schaaf90a}.
The tangential field increases the thermal insulation
of the envelope, because the Larmor rotation of electrons
reduces the transverse electron thermal conductivity.

The case of arbitrary angle $\theta_B$ between the field lines
and the normal to 
the surface was studied by \citet{gh83} in the approximation
of constant (density and temperature independent)
longitudinal and transverse thermal conductivities.
The authors proposed a simple formula
which expresses $T_\mathrm{s}$ at arbitrary $\theta_B$
through two values of  $T_\mathrm{s}$ 
calculated at $\theta_B=0$ and $90^\circ$.
The case of arbitrary $\theta_B$
was studied also by \citet{schaaf} 
and \citet{hh-theory,hh-multi}.

\citet{PY01} reconsidered the thermal structure
of blanketing iron envelopes for any $\theta_B$, using improved thermal
conductivities \citep{P99}. 
\citet{PYCG03} analyzed 
accreted 
  blanketing
envelopes composed of light elements. 
In agreement with an earlier conjecture
of \citet{hern85} and simplified treatments of \citet{page95} and
\citet{shibyak}, they demonstrated that the dipole magnetic field
(unlike the radial one) does not necessarily increase the total
stellar luminosity $L_\gamma$ at a given $T_\mathrm{b}$. On the
contrary, the field $B\sim10^{11}$--$10^{13}$~G lowers $L_\gamma$,
and only the fields $B\gtrsim10^{14}$ G significantly increase it.
\citet{PYCG03} shifted the inner boundary of the blanketing envelope
to the neutron drip, $\rho_\mathrm{b}=4\times10^{11}$ g~cm$^{-3}$,
because in some cases they found
a non-negligible temperature drop at $\rho>10^{10}$ g~cm$^{-3}$.
They obtained that magnetized accreted envelopes 
are generally more heat-transparent than non-accreted ones
(the same is true in the field-free case, studied by \citealt*{PCY}).
However, this heat transparency enhancement is less significant, when 
the transparency is already enhanced by a superstrong magnetic field.

Recently \citet*{PUC} showed that qualitatively the same 
dependence of $L_\gamma$ on $B$ 
and on the chemical composition holds
not only for dipole, but also for small-scale 
field configurations.

\citet*{GKP04,GKP06} studied heat-blanketing envelopes 
with a magnetic field 
anchored either in the core or in the inner crust of the star
(with dipole and toroidal field components of different
strengths). They showed that 
a superstrong field in the inner crust 
of a not too hot star can significantly affect
the surface temperature distribution and make it nonuniform 
and even asymmetric,
with hot spots having different temperatures.
Similar results were obtained by \citet*{perez-azorin06},
who also evaluated pulsed fractions and phase-dependent spectra
of neutron stars with strong magnetic fields anchored in the inner crust.

\subsection{Heat-Blanketing Envelopes of Magnetars:
The Effect of Neutrino Emission}
\label{sect:magnetars}

Magnetars differ from ordinary pulsars in two respects:
they possess superstrong surface magnetic fields,
and they are generally younger and hotter.
The first circumstance suggests to extend the
heat-blanketing layer to deeper 
layers (at least to the neutron drip density
as was done by \citealt{PYCG03}). 
The second 
indicates that neutrino emission 
can be important in the heat-blanketing envelopes.
Accordingly,
we have modified our computer code, which calculates 
the thermal structure \citep{PY01,PYCG03},
to make it fully relativistic and to allow for energy sinks
during heat diffusion.

\begin{figure*}
\centering
  \includegraphics[width=0.75\textwidth]{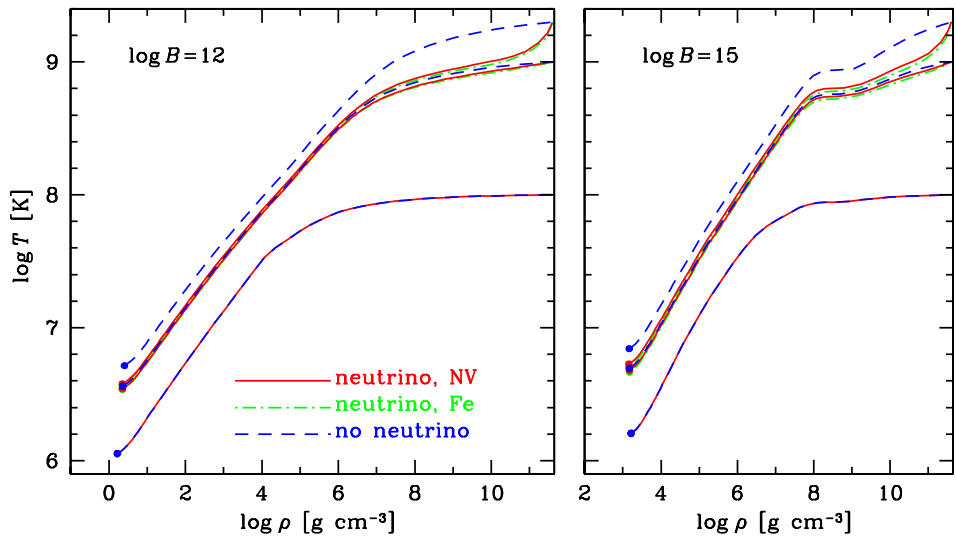}
\caption{Temperature profiles in the outer crust of a neutron star
with magnetic field $B=10^{12}$~G (left panel) or $B=10^{15}$~G
(right panel), directed perpendicular to the stellar surface
($\theta_B=0$).
Solid lines (NV) -- ground-state (Negele--Vautherin)
nuclear composition of the envelope,
dot-dashed lines -- $^{56}$Fe envelope. 
For comparison, dashed lines show the temperature profiles
without allowance for neutrino emission (for ground-state matter).
For every family of curves, temperature at the neutron drip density
is fixed to $T_\mathrm{b}=10^8$~K, $10^9$~K, and $2\times10^9$~K.
The dots at the left end of the profiles correspond to
the radiative surface, where the optical depth equals 2/3 and
$\sigma T^4 = F_R$.
}
\label{fig:T1}
\end{figure*}

\subsubsection{Basic Equations}

A complete set of equations for mechanical and thermal
structure of a spherically symmetric star in hydrostatic
equilibrium has been derived 
by \citet{Thorne77}.
They can easily be transformed to the form valid in an
envelope of a star with radial heat transport, anisotropic (slowly
varying) temperature distribution over any spherical layer,
and a force-free magnetic field.  
Assuming quasistationary heat transport
and neutrino emission,
these equations reduce to the following system
of ordinary differential equations
for the metric function (gravitational potential) $\Phi$,
the local heat flux $F_r$,
temperature $T$,
and gravitational mass $m$, contained within a sphere of 
circumferential radius $r$,
as functions of pressure $P$:
\begin{eqnarray}&&
   \frac{\mathrm{d}\Phi}{\mathrm{d}\ln P} = -\frac{1}{\mathcal{K}_h}
   \frac{P}{\rho c^2},
\label{Phi}
\\&&
   \frac{1}{r^2}\,\frac{\mathrm{d} (r^2 F_r)}{\mathrm{d}\ln P} =
     \frac{P}{\rho g} 
     \frac{Q}{\mathcal{K}_h \mathcal{K}_g}
     - 2 F_r \frac{\mathrm{d}\Phi}{\mathrm{d}\ln P},
\label{L_r}
\\&&
   \frac{\mathrm{d}\ln T}{\mathrm{d}\ln P} =
     \frac{3}{16} \frac{F_r}{\sigma T^4}
     \frac{K'P}{g} \frac{1}{\mathcal{K}_h \mathcal{K}_g}
      - \frac{\mathrm{d}\Phi}{\mathrm{d}\ln P},
\label{T}
\\&&
   \frac{\mathrm{d} r}{\mathrm{d}\ln P} =
     - \frac{P}{\rho g} \frac{\mathcal{K}_r }{\mathcal{K}_h \mathcal{K}_g}, 
\\&&
   \frac{\mathrm{d} m}{\mathrm{d}\ln P} =
     - \frac{4\pi r^2 P \mathcal{K}_r }{g \mathcal{K}_h \mathcal{K}_g}. 
\label{m}
\end{eqnarray}
Here $\rho$ is the mass density (equivalent energy density of the matter),
$Q$ is the net energy loss per unit volume
($Q=Q_\nu$ is the neutrino 
emissivity in our case, although
generally $Q=Q_\nu-Q_\mathrm{h}$,
$Q_\mathrm{h}$ being the heat deposition rate,
e.g., due to 
nuclear reactions),
$K'=K\rho_\mathrm{bar}/\rho$,
$K$ is the opacity, 
$\rho_\mathrm{bar}=n_\mathrm{bar} m_\mathrm{H}$ 
is the so-called ``baryon mass density,''
$n_\mathrm{bar}$ is the baryon number density,
$m_\mathrm{H}$ is the mass of the hydrogen atom,
$\sigma$ is the Stefan-Boltzmann constant,
$g=Gm/(r^2\mathcal{K}_r)$ is the local gravity, and
$G$ is the gravitational constant.
Furthermore,
\begin{eqnarray}&&
   \mathcal{K}_r = (1-2Gm/rc^2)^{1/2},
\label{K_r}
\\&&
   \mathcal{K}_h = 1 + P/\rho c^2,
\\&&
   \mathcal{K}_g = 1 + 4\pi r^3 P/mc^2,
\label{K_g}
\end{eqnarray}
are general relativistic corrections to radius, enthalpy and
surface gravity, respectively.
We adopt the conventional definition of the opacity $K$,
used also by \citet{Thorne77}.
In our notations $K'=16\sigma T^3/(3\kappa \rho)$,
where $\kappa$ is the total (electron plus radiative)
thermal conductivity of the plasma.
Note that $\rho\approx\rho_\mathrm{bar}$ and $K'\approx K$
in the entire neutron star envelope.
We use the same thermal conductivities as in \citet{PY01}.
The effective radial thermal conductivity in a local part
of the surface equals
$\kappa = \kappa_\| \cos^2\theta_B + \kappa_\perp \sin^2\theta_B$,
where $\kappa_\|$ and $\kappa_\perp$ are the components of the
conductivity tensor responsible for heat transport along and across
field lines, respectively.

The local (non-neutrino) luminosity equals the integral of 
the flux over the sphere of radius $r$,
\begin{equation}
   L_r = \int\sin\theta\mathrm{d}\theta\,\mathrm{d}\varphi\,
    r^2 F_r(\theta,\varphi),
\end{equation}
where $(\theta,\varphi)$ are the polar and azimuthal angles.
For a magnetic dipole model \citep{GinzburgOzernoy}, 
$\tan\theta= 2\tan\theta_B$.

The boundary conditions to Eqs.~(\ref{Phi})\,--\,(\ref{m}) 
are
$$ 
   \Phi_\mathrm{s} = \ln\mathcal{K}_{r\mathrm{s}}, 
~ 
   F_{r\mathrm{s}} = F_R=\sigma T_\mathrm{s}^4 ,
~ 
   r_\mathrm{s} = R,
~ 
   m_\mathrm{s} = M,
$$ 
and the surface pressure is determined, following Gud\-mundsson, Pethick, \& Epstein (\citeyear{GPE}),
by the condition $K_\mathrm{rad,s} P_\mathrm{s}/g_\mathrm{s}$ $=2/3$,
where $K_\mathrm{rad}$ is the radiative opacity.
The subscript `s' refers to surface values; 
$M$ is the gravitational stellar mass.

\begin{figure}
\centering
  \includegraphics[width=0.37\textwidth]{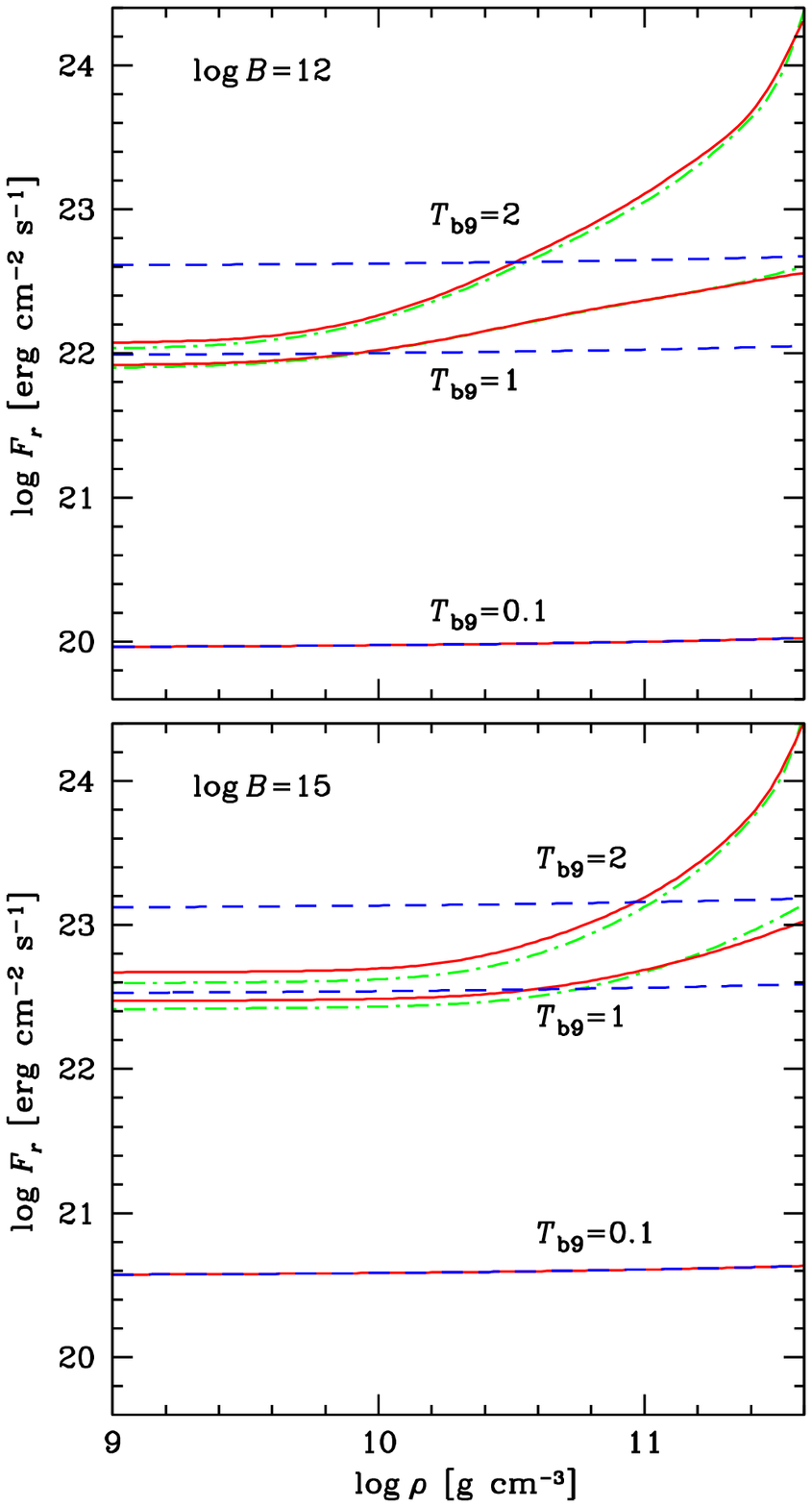}
\caption{Profiles of local radial heat flux $F_r$ for the cases shown
in Fig.~\ref{fig:T1}. 
Top panel: $B=10^{12}$~G,
bottom panel: $B=10^{15}$~G; solid lines -- ground-state matter,
dot-dashed lines -- $^{56}$Fe, dashed lines -- without neutrino emission
 for ground-state matter.}
\label{fig:F}
\end{figure}

General Relativity correction factors $\mathcal{K}_r$, $\mathcal{K}_h$,
and $\mathcal{K}_g$ in Eqs.\ (\ref{K_r})\,--\,(\ref{K_g})
are nearly constant because
$(M-m)/M\sim10^{-5}$ and
$P/\rho c^2\sim10^{-2}$ at the bottom of the outer crust.
However, they are taken into account in our code, in order
to extend calculations to deeper neutron star layers, when required.

Equations (\ref{L_r}), (\ref{T}) are one-dimensional,
which implies that the mean temperature gradient along stellar radius
is large compared to the tangential temperature gradient, i.e.,
$\epsilon\equiv|\partial T/\partial x|/|\partial T/\partial r|\ll1$,
where $x$ is a coordinate along the stellar surface.
Let us roughly estimate the mean value of $\epsilon$
for a large-scale (e.g., dipole) magnetic field,
following \citet{gh83}. In this case,
$\epsilon\sim (\bar{T}_\mathrm{s}/T_0) (l_0/R)$, 
where $l_0$ is the depth at which the temperature
distribution becomes nearly isotropic ($T=T_0$ at the depth $l_0$)
and $\bar{T}_\mathrm{s}$ is the mean surface temperature.
At the bottom of the outer crust 
(i.e., assuming $T_0=T_\mathrm{b}$)
we have $l_0/R\lesssim0.1$ (e.g., $l_0\approx0.6$ km
for $M=1.4 M_\odot$ and $R=10$ km)
and $\bar{T}_\mathrm{s}/T_0\sim 10^{-2}$ (see Sect.~\ref{sect:res}).
This estimate gives $\epsilon\lesssim 10^{-3}$.

Nevertheless, the one-dimensional approximation
is inaccurate at those loci where magnetic field
lines are tangential to the surface, because in this case 
one cannot neglect
their curvature, $\partial\theta_B/\partial x$.
For large-scale magnetic field
($\partial\theta_B/\partial x\sim R^{-1}$),
the maximum size $a$ of such sites can be estimated as the distance
at which an initially tangential field line crosses
the depth $l_0$ (which assumes that heat flows along field lines,
i.e., $\kappa_\perp/\kappa_\|\ll l_0/a$).
Thus, $a\lesssim\sqrt{R l_0}$. 
Since ${T}_\mathrm{s}$ is minimal at such sites,
their contribution to the total stellar luminosity
can be neglected in the first approximation.

\begin{figure}
\centering
  \includegraphics[width=0.48\textwidth]{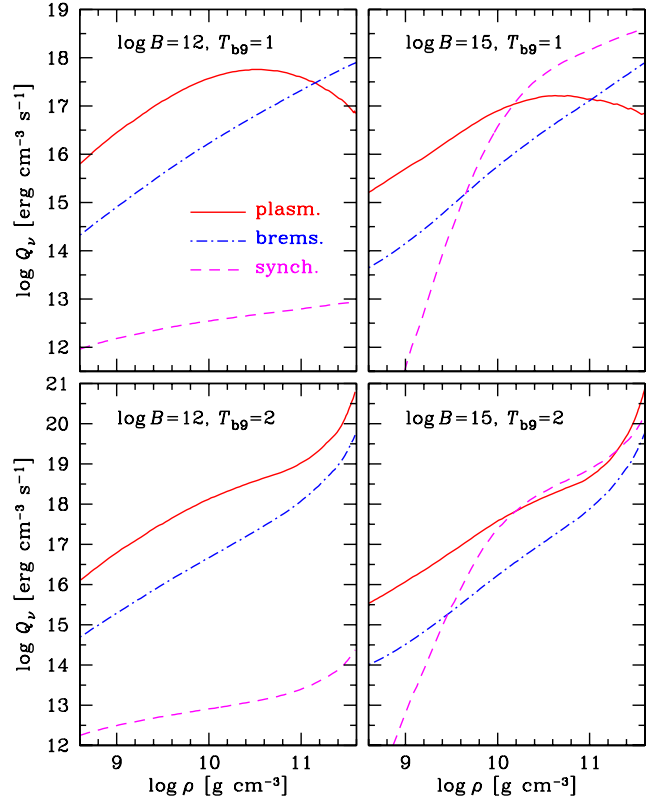}
\caption{Contributions to the total neutrino emissivity $Q_\nu$
from plasmon decay (solid lines), electron neutrino bremsstrahlung
(dot-dashed lines), and electron synchrotron radiation (dashed lines).
Left panels: $B=10^{12}$~G (radial field); right panels: $B=10^{15}$~G; 
top panels: $T=10^9$~K; bottom panels: $T=2\times10^9$~K. 
}
\label{fig:Q}
\end{figure}

\subsubsection{Results}
\label{sect:res}

We have solved Eqs.~(\ref{Phi})\,--\,(\ref{m}) using a 
straightforward generalization of the Runge-Kutta method
employed in our previous papers (\citealt{PCY}; 
Potekhin \& Yakovlev \citeyear{PY01};
\citealt{PYCG03}).
Temperature profiles have been calculated within a local
part of the blanketing envelope with a locally constant
magnetic field.
At every value of $P$, a corresponding value of $\rho$ 
was found from the equation of state of 
magnetized electron-ion relativistic plasma (e.g., \citealt{PY96}),
using approximations of Fermi-Dirac integrals presented in \citet{PC00}.
In all examples shown in the figures we chose 
the ``canonical'' neutron star
mass $M=1.4 M_\odot$ and radius $R=10$ km.
The neutrino emissivity is calculated as
$Q_\nu = Q_\mathrm{pair} +  Q_\mathrm{pl}+ Q_\mathrm{syn}
 + Q_\mathrm{brems}$,
where the contributions due to electron-positron pair annihilation 
$Q_\mathrm{pair}$, plasmon decay $Q_\mathrm{pl}$,
synchrotron radiation of neutrino pairs by electrons $Q_\mathrm{syn}$,
and
bremsstrahlung in electron-nucleus collisions $Q_\mathrm{brems}$
are given, respectively, by Eqs.~(22), (38), (56), and (76)
of \citet{YKGH}. According to the results of  \citet{Itoh-ea} and \citet{YKGH},
other neutrino emission mechanisms are unimportant
in the outer crust of neutron stars.

Our calculations
show that neutrino emission is crucially important
for the thermal structure of neutron stars with internal
temperature $T_\mathrm{b}\gtrsim10^9$~K.

Figure \ref{fig:T1} shows temperature profiles at $T_\mathrm{b}=10^8$~K,
$10^9$~K, and $2\times10^9$~K for magnetic fields
$B=10^{12}$~G and $10^{15}$~G, directed perpendicular
to the stellar surface. The present results (solid and 
dot-dashed lines) are compared with 
the profiles calculated neglecting neutrino emission (dashed lines).
At the lowest temperature $T_\mathrm{b}=10^8$~K
there is virtually no difference (all the lines coincide).
At $T_\mathrm{b}=10^9$~K, the difference is noticeable,
and at $T_\mathrm{b}=2\times10^9$~K it is large.
Because of the growth of neutrino emission with increasing
temperature at $\rho\gtrsim10^{10}$~ g~cm$^{-3}$,
$T_\mathrm{s}$ is nearly independent of
$T_\mathrm{b}$ at $T_\mathrm{b}\gtrsim 10^9$~K,
but depends on the magnetic field.

In this figure we also compare 
temperature profiles for the different heavy-element
compositions of the outer envelope: 
iron (dot-dashed lines) and ground-state
matter \citep{NV73}. The effect of composition is not strong,
but noticeable when the neutrino emission is important,
because $Q_\nu$ depends on the electron number density 
that is a function of composition for a given $\rho$.

Figure \ref{fig:F} demonstrates the profile of the local heat flux $F_r$,
for the same cases as in Fig.~\ref{fig:T1}, plotted by the same
line types. Without neutrino emission (dashed lines),
$F_r$ would be nearly constant, with only $\approx2$\% increase
towards the inner crust due to the General Relativity effects
(associated with the variation of 
the metric function $\Phi$) and $\approx9$\% increase
because of the spherical geometry (the $r^2$ factor). 
The neutrino emission leads to a strong dependence of the flux
on $\rho$ and violates the familiar relation between
$F_r$ and $T_\mathrm{s}$ derived in the absence of
energy sinks.

Figure \ref{fig:Q} shows which neutrino emission
mechanism dominates at given $\rho$, $T_\mathrm{b}$, and $B$.
For superstrong magnetic fields, the neutrino synchrotron mechanism
dominates in certain density ranges, which does not happen
at ``ordinary pulsar'' $B \sim 10^{12}$~G. 
It is natural because of the strong $B$-dependence of the synchrotron
neutrino emissivity.
Pair annihilation neutrino emissivity is not seen in the figure,
because it is too small.
Notice that the emissivity of plasmon decay and
bremsstrahlung processes can be affected by superstrong
magnetic fields which has not been explored so far
(and we present the emissivities in the field-free case).
A slow dependence of these emissivities on $B$,
seen in Fig.~\ref{fig:Q}, is indirect (caused by
the dependence of temperature profiles on $B$).

\begin{figure}
\centering
  \includegraphics[width=0.4\textwidth]{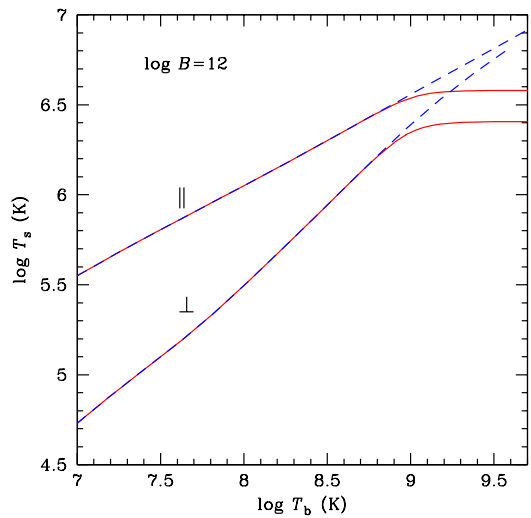}
\caption{Surface temperature $T_\mathrm{s}$ as a function of 
temperature $T_\mathrm{b}$ at the neutron drip point
for a neutron star with the magnetic field $B=10^{12}$~G, 
directed along stellar radius
($\theta_B=0$, parallel heat transport, sign $\|$) or along the surface
($\theta_B=90^\circ$, transverse transport, $\perp$).
Ground-state composition is assumed.
Solid lines -- present calculation,
dashed lines -- neutrino emission is neglected.}
\label{fig:pfs12}
\end{figure}

\begin{figure}
\centering
  \includegraphics[width=0.4\textwidth]{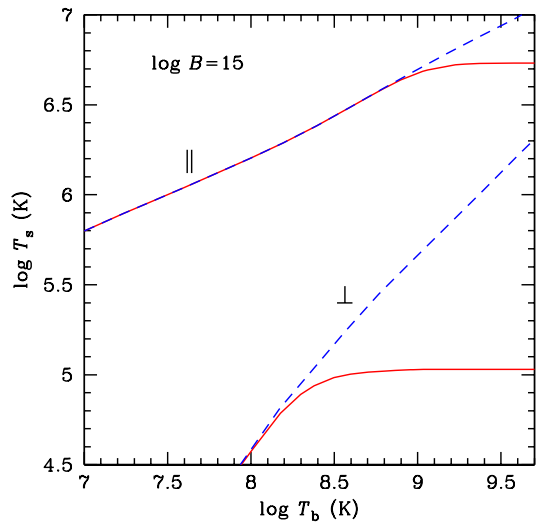}
\caption{The same as in Fig.~\ref{fig:pfs12},
but for $B=10^{15}$~G.}
\label{fig:pfs15}
\end{figure}

Figures~\ref{fig:pfs12} and \ref{fig:pfs15} give the
$T_\mathrm{s}(T_\mathrm{b})$ relation
for the magnetic fields $B=10^{12}$~G and $B=10^{15}$~G
perpendicular and parallel to the radial direction. 
The relations obtained with and without allowance for neutrino emission
are plotted by solid and dashed lines, respectively.
We see
that at $T_\mathrm{b}\lesssim10^8$~K the neutrino emission does not
affect $T_\mathrm{s}$. At higher $T_\mathrm{b}\gtrsim10^9$~K, in contrast,
this emission is crucial: if $Q_\nu=0$,
then $T_\mathrm{s}$ continues to grow up with increasing $T_\mathrm{b}$,
whereas with realistic $Q_\nu$ the surface temperature tends to 
a constant limit, which depends on $\theta_B$ and $B$.
In most cases this limit is reached when $T_\mathrm{b}\sim10^9$~K,
but for a superstrong field (right panel) and transverse
heat propagation, it is reached at still smaller 
$T_\mathrm{b}\sim3\times10^8$~K.
%

\begin{figure*}
\centering
  \includegraphics[width=0.65\textwidth]{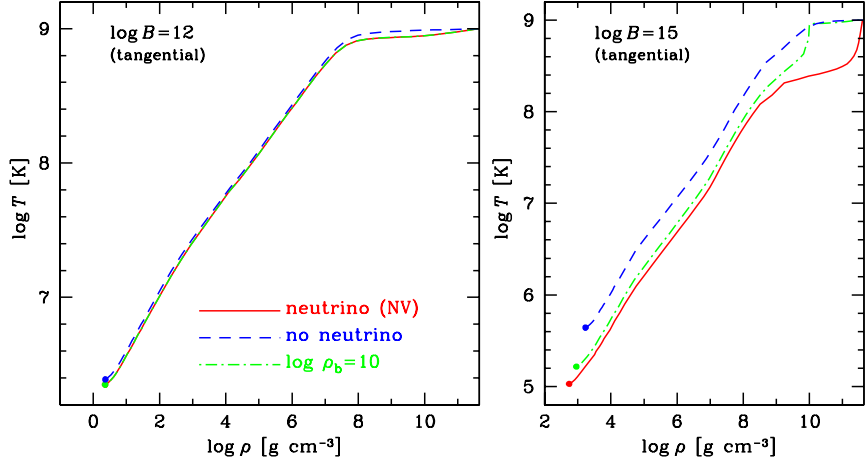}
\caption{Temperature profiles in the outer crust of a neutron star
with the magnetic field $B=10^{12}$~G (left panel) or $B=10^{15}$~G
(right panel), directed parallel to the stellar surface,
with the temperature $T_\mathrm{b}=10^9$~K at neutron drip density 
and the ground-state composition of the matter.
Solid lines -- present calculation,
dot-dashed lines -- with a switch to non-magnetic ($B=0$) calculation
at $\rho>10^{10}$ g~cm$^{-3}$,
dashed lines -- without neutrino energy losses.}
\label{fig:T0}
\end{figure*}

In Fig.~\ref{fig:T0} we explore joint effects
of magnetic field, neutrino emission,
and the shift of the inner boundary from $10^{10}$ g~cm$^{-3}$
to the neutron drip. Here magnetic field lines are directed
along the surface (perpendicular to the direction of heat
transport). Therefore, these effects are most pronounced.
The solid lines are the results of accurate calculations;
the dashed lines, as before, show the model with $Q_\nu=0$;
and dot-dashed lines
are obtained by solving Eqs.~(\ref{Phi})\,--\,(\ref{m})
in the whole domain $\rho_\mathrm{s}\leq\rho\leq 4\times10^{11}$
g~cm$^{-3}$, but with the magnetic field artificially ``switched off''
at $\rho>10^{10}$ g~cm$^{-3}$. 
This is a simulation of the model, where the heat transport in the magnetized
plasma is solved accurately up to
$\rho_\mathrm{b}=10^{10}$ g~cm$^{-3}$,
while after this boundary nonmagnetic 
heat balance and transport equations are solved.
In the absence of the results reported here,
the latter model was used by \citet{Kam06}
to study the thermal structure and evolution of magnetars 
as cooling neutron stars 
with a phenomenological heat source in a spherical internal layer.
In the left panel of Fig.~\ref{fig:T0}, the solid and dot-dashed lines
almost coincide, indicating that in this case 
$\rho_\mathrm{b}=10^{10}$ g~cm$^{-3}$ may provide a sufficient accuracy.
In the right panel, in contrast, the effect of the shift of the
inner boundary is quite visible.
Therefore, we conclude that
the development of our thermal structure code, reported here,
will allow us to study the thermal history of magnetars
at a higher accuracy level.

Meanwhile, a comparison of the profiles shown in Figs.~\ref{fig:T1}
and \ref{fig:T0} prompts that at $T\gtrsim10^9$~K and 
$B\sim10^{15}$~G the magnetic effects on the conductivity
could be important at still higher densities in the inner crust.
Investigation of this possibility requires taking
into account the effects of free (possibly superfluid)
neutrons on thermal conduction and neutrino emission.
We are planning to perform such study in the future.

\begin{figure}
\centering
  \includegraphics[width=0.48\textwidth]{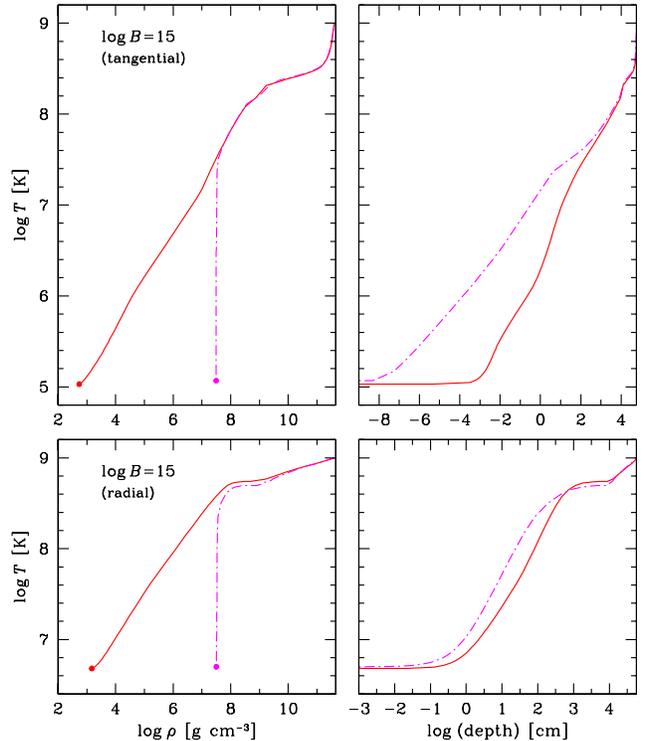}
\caption{Temperature in the outer crust of a neutron star
as a function of density (left panels) or proper depth
behind the radiative surface (right panels).
Magnetic field $B=10^{15}$~G is
directed parallel (top panels) or perpendicular (bottom panels)
to the stellar surface; $T_\mathrm{b}=10^9$~K.
Solid lines -- the model of diffuse surface (no phase transition),
dot-dashed lines -- condensed surface.}
\label{fig:pfc}
\end{figure}

In the above figures we have shown the results of calculations
where the surface of a neutron star was assumed diffuse, i.e.,
without the phase transition 
discussed in Sect.~\ref{sect:surf}. However, for $B=10^{15}$~G
and $T_\mathrm{b}\lesssim10^9$~K the surface temperature
$T_\mathrm{s}$ is below the estimates of the critical temperature $T_c$
given in Sect.~\ref{sect:surf}.
This possibility is explored in Fig.~\ref{fig:pfc}.
Here we use the equation of state containing nonideal
terms for strongly magnetized fully ionized plasma
(Sect.~IIIB of \citealt{PCS}), which enforce 
the phase transition.
At $B=10^{15}$~G the surface density is
very high,
$\rho_\mathrm{cond}\approx3.1\times10^7$ g~cm$^{-3}$,
in our model
(and the analytic estimate in Sect.~\ref{sect:surf} gives
$\rho_\mathrm{cond}\sim2\times10^7$ g~cm$^{-3}$).
The solid lines in the left panels
reproduce the profiles shown in Figs.~\ref{fig:T1}
and~\ref{fig:T0}, whereas the dot-dashed lines 
display the case of magnetic condensation. In the right panels,
we show the same temperature profiles as a function of local
proper depth $l$ ($\mathrm{d}l = - \mathcal{K}_r^{-1} \mathrm{d} r$),
measured from the radiative surface. Although the temperature profiles
with and without magnetic condensation are drastically different
in the surface layers, the effective temperature remains almost the same.
Thus, the magnetic condensation does not 
significantly affect the $T_\mathrm{s}(T_\mathrm{b})$
relation.
This should not be surprising, because, as explained by \citet{GPE},
the main regulator of the $T_\mathrm{b}$\,--\,$T_\mathrm{s}$
relation is the ``sensitivity strip'' where $\kappa$
has a minimum. This domain, a real bottleneck
for the heat outflow, lies at $\rho>\rho_\mathrm{cond}$
(except for low $T_\mathrm{s}$, not considered here),
and therefore it is not affected by the condensation.

\section{Summary}

We have described the main effects of strong magnetic fields on
the properties of neutron star atmospheres and heat blanketing envelopes,
with the emphasis on the difference between ordinary neutron stars
and magnetars. 
Observations of magnetars pose new theoretical problems and challenges
because of magnetars' superstrong magnetic fields and high temperatures.
We also report a solution to one of such problems, which consists
in taking into account neutrino energy losses
in the outer crust of hot, strongly magnetized neutron stars.
We have demonstrated that, because of these losses,
 the effective surface temperature $T_\mathrm{s}$ 
 almost ceases to depend
 on the temperature $T_\mathrm{b}$ in the inner crust
  as soon as  $T_\mathrm{b}$ exceeds $10^9$~K.
A direct consequence of this observation is that
in the absence of powerful energy sources in outer envelopes,
the stationary (time-averaged) 
effective temperature cannot be raised above
the value that it would have at $T_\mathrm{b}\approx10^9$~K,
irrespectively of the energy release in the deeper layers.

\begin{acknowledgements}
We are grateful to the anonymous referee for useful comments.
We thank A.D.\,Kaminker for pointing out to some misprints
in formulae, which are corrected in this electronic version
of the paper.
AYP thanks the organizers of the conference ``Isolated Neutron
Stars: from the Interior to the Surface'' (London, April 24--28, 2006), 
especially
Silvia Zane and Roberto Turolla, for perfect organization,
attention and support. Fruitful discussions
with participants of this conference
have significantly contributed to the developments
partly reported in the present paper.
\end{acknowledgements}

\end{document}